\documentclass[twocolumn]{aastex631}

\usepackage{booktabs}
\usepackage{tabularx}
\usepackage{stix}
\usepackage{amsmath}
\usepackage{fontawesome5}
\usepackage{xcolor}
\usepackage{float}
\usepackage{longtable}
\usepackage{threeparttable}
\usepackage{afterpage}
\usepackage{footnote}

\begin{document}

\title{Out of the Darkness: High-resolution Detection of CO Absorption on the Nightside of WASP-33b}

\author[0009-0001-6922-1261]{Georgia Mraz}
\affiliation{Department of Physics and Trottier Space Institute, McGill University, 3600 rue University, Montréal, Québec, H3A 2T8, Canada}

\author[0000-0002-7786-0661]{Antoine Darveau-Bernier}
\affiliation{Department of Physics and Institute for Research on Exoplanets, 1375 Ave.\ Thérèse-Lavoie-Roux, Montréal, QC H2V 0B3, Canada} 

\author[0000-0001-9427-1642]{Anne Boucher}
\affiliation{Department of Physics and Trottier Space Institute, McGill University, 3600 rue University, Montréal, Québec, H3A 2T8, Canada}
\affiliation{Department of Physics and Institute for Research on Exoplanets, 1375 Ave.\ Thérèse-Lavoie-Roux, Montréal, QC H2V 0B3, Canada}  

\author[0000-0001-6129-5699]{Nicolas B.\ Cowan}
\affiliation{Department of Physics and Trottier Space Institute, McGill University, 3600 rue University, Montréal, Québec, H3A 2T8, Canada}
\affiliation{Department of Earth \& Planetary Sciences, McGill University, 3450 rue University, Montr\'eal, QC, H3A 0E8, Canada}

\author[0000-0002-6780-4252]{David Lafreni\`ere}
\affiliation{Department of Physics and Institute for Research on Exoplanets, 1375 Ave.\ Thérèse-Lavoie-Roux, Montréal, QC H2V 0B3, Canada} 

\author[0000-0001-9291-5555]{Charles Cadieux}
\affiliation{Department of Physics and Institute for Research on Exoplanets, 1375 Ave.\ Thérèse-Lavoie-Roux, Montréal, QC H2V 0B3, Canada}



\begin{abstract}

We observed the ultra hot Jupiter WASP-33b with the Spectro-Polarimètre Infra-Rouge on the Canada Fance Hawaii Telescope. 
Previous observations of the dayside of WASP-33b show evidence of CO and Fe emission indicative of a thermal inversion. We observed its nightside over five Earth-nights to search for spectral signatures of CO in the planet's thermal emission. Our three pre-transit observations and two post-transit observations are sensitive to regions near the morning or evening terminators, respectively. From spectral retrievals, we detect CO molecular absorption in the planet's emission spectrum after transit at $\sim$6.6$\sigma$. This is the strongest ground-based detection of nightside thermal emission from an exoplanet, and only the third ever. CO appearing in absorption suggests that the nightside near the evening terminator does not have a temperature inversion; this makes sense if the dayside inversion is driven by absorption of stellar radiation. On the contrary, we do not detect CO from the morning terminator. This may be consistent with heat advection by an eastward jet. Phase-resolved high-resolution spectroscopy offers an economical alternative to space-based full-orbit spectroscopic phase curves for studying the vertical and horizontal atmospheric temperature profiles of short-period exoplanets.
\end{abstract}



\section{Introduction} \label{sec:intro}

Ultra-hot Jupiters (UHJ) are characterized by short orbital periods of less than two days and can receive over 100 times the stellar irradiation of their Hot Jupiter counterparts \citep{baxter_transition_2020}. As for all short-period planets, synchronous rotation can cause extreme temperature differences between their day and nightsides \citep{showman_atmospheric_2002}, exceeding 1000 K for UHJs \citep{cowan_thermal_2012} leading to molecular dissociation on the dayside \citep{bell_very_2017} and recombination on the nightside \citep{bell_increased_2018}.

The extreme stellar irradiation experienced by UHJs causes dayside thermal inversions: the atmospheric temperature increases with decreasing pressure \citep{lothringer_influence_2019,beltz_exploring_2022}. This is due to the presence of optical or ultraviolet absorbers in the upper atmosphere which absorb irradiation from the hot star (\citealt{fortney_unified_2008}). The nightside, however, is not irradiated, and thus, should not have an inversion. For an inversion to occur on the nightside of the planet, the time scale for day-to-night heat transport at the high altitudes of the inversion would need to be shorter than the timescale for those upper layers to radiate away their heat. As shown by current general circulation models (GCM), it is likely that the nightside will present a non-inverted atmosphere due to the short radiative timescales in the upper layers \citep{van_sluijs_carbon_2023, gandhi_coupled_2020}.

Using a full orbit spectroscopic phase curve from the Hubble Space Telescope's WCF3/G141, \cite{mikal-evans_diurnal_2022} showed that the dayside of UHJ WASP-121b exhibits water in emission while its nightside has water absorption features. Probing the vertical temperature profile of both a planet's day- and nightside typically requires full-orbit space-based spectroscopic phase curves, which are resource-intensive. 

In principle, similar 2D information about an exoplanet's atmosphere could be obtained via ground-based high-resolution spectroscopy (HRS). Although continuous observations spanning a full exoplanetary orbit are impossible, the planet's changing Doppler signature makes it possible to disentangle the exoplanetary signal from that of its host star or the Earth's own atmosphere. Observing a planet just before and after transit would allow for a glimpse of the thermal emission from the nightside of the planet without the confounding factor of transmission features due to the transit of the planet in front of its star.

Although the dayside emission of short-period exoplanets has been detected numerous times via HRS \citep[e.g.,][]{finnerty_keckkpic_2023,yan_detection_2021,herman_dayside_2022,cont_atmospheric_2022}, there have only been two detections of nightside emission \citep{matthews_doppler_2024,yang_evidence_2024}. One tentative ground-based detection of CO on the nightside of HD 179949b was accomplished using Doppler tomography, similar to cross- correlataion \citep{watson_doppler_2019}, with the Cryogenic High-Resolution Infrared Echelle Spectrograph on the Very Large Telescope in Cerro Paranal \citep{matthews_doppler_2024}. \cite{yang_evidence_2024} found evidence of H$_2$O emission on the nightside indicating an inverted temperature structure using data from CARMENES and GIANO-B.

Another prime candidate for high-resolution emission spectroscopy is WASP-33b \citep{collier_cameron_line-profile_2010}, as it is among the most highly irradiated exoplanets with a period of only 1.2 days around an A5 star \citep{smith_thermal_2011} and an inflated radius of 1.6 Jupiter radii ($R_{\rm Jup}$). 
Table \ref{tab:wasp33_parameters} lists the full star and planet parameters. Previous high-resolution transit observations of WASP-33b in the optical have revealed detections such as Ionized Calcium absorption\citep{yan_ionized_2019}, H$\alpha$ absorption and H$\beta$ absorption \citep{borsa_gaps_2021,yan_detection_2021,yan_detection_2022, cauley_time-resolved_2021}. High-resolution near-infrared emission spectroscopy detected Ti emission, V emission, TiO emission \citep{Nugroho_wasp33} \citep{cont_atmospheric_2022}, and tentative FeH absorption\citep{kesseli_search_2020}. Moreover, CO, Fe, OH, and H$_2$O emission have been detected on dayside thermal emission, indicating a temperature inversion \citep{herman_day-side_2022,van_sluijs_carbon_2023,finnerty_keckkpic_2023}. Non-detections of TiO, Ti, VO, and V have also been reported in high-resolution transit spectroscopy \citep{yang_high-resolution_2023}.

Few attempts to observe the nightside of WASP-33b have been made. \cite{herman_search_2020} searched for TiO on the nightside of the planet and found no evidence while \cite{yang_evidence_2024} found evidence of a thermally inverted nightside atmosphere when water was found in emission. However, analyzing nightside emission in transit observations remains challenging as the signal needs to be disentangled from the transmission signal as it shows different altitudes in the atmosphere than seen in emission spectroscopy.

As an alternative which does not suffer from this bias a planet can be, observed before and after transit. Contrary to previous studies, we exclude all in-transit parts of the data to focus on the planet's thermal emission and hence vertical atmospheric temperature profile. Here, a non-inverted atmosphere will yield an absorption feature, while an inverted atmosphere will yield an emission feature. An isothermal atmosphere will yield a non-detection. 

We present the nightside detections of CO made from the ground via high-resolution spectroscopy and produce the first nightside temperature profile for WASP-33b. The observations are described in Section \ref{sec:obs}. We describe how we used the reduction and analysis pipeline \texttt{STARSHIPS} in Section~\ref{sec:Reduction}. We show our results in Section~\ref{sec:Results}, discuss them in Section \ref{sec:discussions}, and conclude in Section \ref{sec:conclusions}.

\begin{table}[htbp]
    \centering
    \begin{threeparttable}
    \begin{tabular}{@{}ll@{}}
        
        \toprule
        \textbf{WASP-33 Parameters} & \\
        \midrule
        Stellar Mass ($M_{\odot} $) & 1.495 $\pm$ 0.031\\
        Stellar Radius ($R_{\odot}$) & 1.444 $\pm$ 0.034\\
        Effective Temperature (K) & 7430 $\pm$ 100 \\
        
        \midrule
        \textbf{WASP-33b Parameters} & \\
        \midrule
        Planet Mass ($M_{\rm Jup}$) & 2.093 $\pm$ 0.139\\
        Planet Radius ($R_{\rm Jup}$) & 1.593 $\pm$ 0.074\\
        Semi-Major Axis (AU) & 0.024 $\pm$ 0.007 \\
        Orbital Period (days) & 1.220 $\pm$ 0.000 \\
        Equilibrium Temperature (K) & 2781 $\pm$ 41\\
        Transit Duration (hours) & 2.851 $\pm$ 0.012\\
        Systemic Velocity (km/s) & -2.700  $\pm$ 0.330\\
        \bottomrule
    \end{tabular}

    \caption{Stellar and planetary parameters of WASP-33b. All parameters are from \cite{chakrabarty_precise_2019} except for systemic velocity, which was taken from the Gaia Archive \tnote{a}.}
    \label{tab:wasp33_parameters}
    \begin{tablenotes}
      \item[a] \url{https://www.cosmos.esa.int/gaia}
    \end{tablenotes}
    \end{threeparttable}

\end{table}

\section{Observations}\label{sec:obs}

\begin{figure}[!tb]
    \centering
    \includegraphics[width=1\columnwidth]{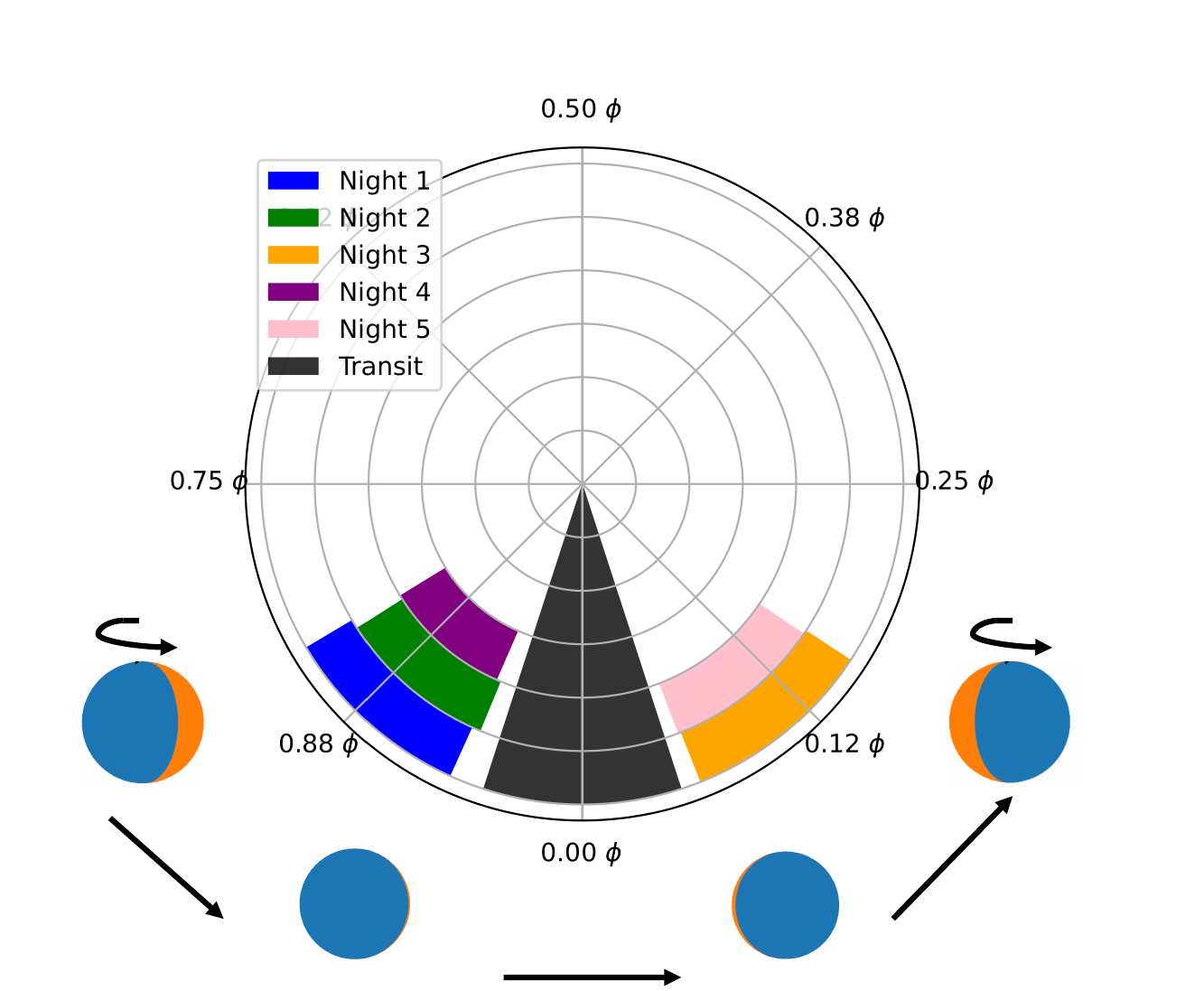} 
    \caption{The orbital phases of our five nightside observations. The black wedge shows the transit of WASP-33b, including a buffer (total shown about $\sim$3 hours). The planet orbits counter-clockwise. Nights 1, 2 and 4 were pre-transit and hence sensitive to regions near the morning terminator, while Nights 3 and 5 were post-transit and sensitive to the evening terminator. Specifically,  Night 1 (2023-09-22) covered phases 0.835--0.932, Night 2 (2023-10-03) covered phases 0.840--0.938, Night 3 (2023-10-29) covered phases 0.059--0.155, Night 4 (2023-11-22) covered phases 0.837--0.935, and Night 5 (2023-11-26) covered phases 0.058--0.155. planet diagrams show orbital phases 0.84, 0.95, 0.06 and 0.15; dayside is shown in orange, and nightside in blue. Arrows indicate orbital motion.}
    \label{fig:obs_time}
\end{figure}

We observed WASP-33b over 5 nights (Program 23BC29, PI A.\ Boucher)  using SPIRou, a near-infrared échelle spectro-polarimeter installed on the 3.6-m Canada-France-Hawaii Telescope (CFHT). The average exposure time for all observations is 183.873 seconds, and each night had 51 exposures. As shown in Figure \ref{fig:obs_time}, three of the five observations were taken pre-transit, while the other two were taken post-transit. These phases in the planet's orbit are sensitive to different regions of the planet. Systematics for observations are shown in Appendix \ref{sec:systematics}.

A synchronously rotating planet rotates in the same direction as its orbital motion. East is defined as the direction of this rotation. Since the predominant winds on hot Jupiters are predicted and observed to be Eastward in most cases, it is useful to distinguish between the evening terminator ($\sim$90$^\circ$ East of the sub-stellar point) where parcels of gas are moving from the dayside to the nightside, and the morning terminator($\sim$90$^\circ$ West of the sub-stellar point), where gas is swept from the nightside to the dayside. Hence, the pre-transit epochs are sensitive to the planet's morning terminator, while the two post-transit epochs probe the planet’s evening terminator. 


SPIRou has a wavelength range of 0.90--2.65 microns and a spectral resolution of $R \approx$ 70,000. All data pre-processing and spectral extraction are performed with A Pipeline to Reduce Observations, \texttt{APERO} version 0.7.288 \citep{cook_apero_2022}. The pixel-to-wavelength calibration uses a combination of simultaneous Fabry-Perot in the reference fiber C and nightly calibration from a hollow-cathode UNe lamp \citep{hobson_spirou_2021}. APERO corrects for diffuse light from the reference fiber that contaminates the science channels A and B. The spectra are extracted individually (A and B), then the flux is combined (AB), resulting in 2D spectra of 49 orders of 4088 (4096 minus 8 reference pixels) pixels each. APERO finally performs a telluric correction that uses a collection of telluric standard calibrators (hot stars) with TAPAS atmospheric models \citep{bertaux_tapas_2014}. Both the extracted 2D spectra and the telluric-corrected spectra are used in the analysis.

\section{Data Analysis } \label{sec:Reduction}

To extract the emission spectra we use the Spectral Transmission and Radiation Search for High Resolution Planet Signal, \texttt{STARSHIPS} \citep{boucher_characterizing_2021,boucher_co_2023}. The pipeline allows for the extraction of the planetary emission signal in the data and cross-correlation with significance testing of model spectra.

Briefly, the process follows five main steps, which are done on an order-by-order basis. 
\begin{enumerate}
    \item Bad pixel correction and masking were applied to the telluric-corrected, blaze-normalized spectra.
    \item The spectra were normalized to the continuum and shifted to the stellar rest frame by applying a Doppler shift to correct for the barycentric Earth radial velocity (BERV).
    \item Deep telluric lines are identified and masked to limit potential telluric contamination. Telluric lines with core transmission below 20\% are considered deep tellurics. The mask is extended from the core of the lines until their 97\% transmission is reached. 
    \item A reference spectrum, built from the aligned spectra in the stellar rest frame, is removed to isolate the emission signal from the planet.
    \item Principal component analysis (PCA) was used to correct for quasi-static residuals from either tellurics or the star. Two principal components were removed.
\end{enumerate}

A visual representation of these steps is shown in Appendix \ref{sec:reduction_details}. Each epoch is reduced separately due to changing atmospheric conditions and nightly calibrations.

\subsection{Planetary Models}

Atmospheric models were then created using \texttt{petitRADTRANS} \citep{molliere_petitradtrans_2019} to produce a synthetic emission spectrum of WASP-33b. The package generates high-resolution spectra which we down-sampled to $R$=250,000. For all models,  H- opacity was also implemented in the continuum \citep{gray_observation_2008}. We create two separate models. The first contains only CO lines with a Guillot temperature profile. Free parameters included atmospheric opacity in the IR wavelengths $\kappa_{IR}$, the ratio of the optical opacities ($\gamma$), and equilibrium temperature $T_{\rm eq}$. The internal temperature was set to $T_{\rm int} = 500\,\mathrm{K}$.

The second model contained Fe \citep{kurucz_including_2018}, H\textsubscript{2}O, CO, OH \citep{rothman_hitemp_2010}, FeH \citep{bernath_mollist_2020},TIO, VO \citep[See references in][]{molliere_petitradtrans_2019}. A Modified Guillot profile was used to allow for more complex line shapes \citep{guillot_radiative_2010,molliere_petitradtrans_2019}. Here, the $ k_{IR}$ is implemented through $\delta\equiv k_{IR}/g$, where $g$ is the gravity of the planet as calculated from parameters in Table \ref{tab:wasp33_parameters}. Other parameters include the pressure where a TP profile deviation can be made ($P_{\rm trans}$) and the factor of intensity of the deviation ($\alpha$). Again $T_{\rm eq}$ was free while $T_{\rm int}$ was set to 500\,K.

\subsection{Detection and Significance}

\texttt{STARSHIPS} performs cross-correlation and other signal significance testing. As detailed in \cite{boucher_characterizing_2021}. Here, we use a similar approach adapted for emission spectroscopy. The cross-correlation function (CCF) used is from \cite{gibson_detection_2020}:
\begin{equation}
\label{eq:CCF}
    CCF(\theta,v_P)=\sum_{i=1}^N{ \frac{f_i m_i(\theta,v_P)}{\sigma_i^2}},
\end{equation}
where $N$ is the number of wavelengths, $f_i$ is the spectrum, $\sigma_i$ is its uncertainty, and $m_i$ is the model. The uncertainty, $\sigma_i$, includes both temporal and spectral variability. The known planetary radial velocity around its host star is $v_P$  and $\theta$ is the model parameter vector. As described in \cite{gibson_detection_2020} and more generally in \cite{brogi_retrieving_2019} it maps the likelihood function:
\begin{equation}
\label{eq:logL}
    \ln \mathcal{L} =-\frac{N}{2}\ln\left( \frac{\chi^2}{N}\right),
\end{equation}
where the badness-of-fit is 
\begin{equation}
\label{eq:chi2}
    \chi^{2}=\sum{\frac{(f_i-m_i)^2}{\sigma_i^2}}.
\end{equation}
We then select the orders in which CO models have the strongest lines, specifically the 46$^{\rm th}$ and 47$^{\rm th}$ orders of our spectra, which are centered at 2.33 and 2.40 microns. 

\texttt{STARSHIPS} also quantifies detection significance with a scaled Welch t-test map (see Figure \ref{fig:combined-results}). This evaluates the null hypothesis that two samples were drawn from the same distribution. The two samples are CCF values close to and far from the planet RV. The CCFs are combined in time for different RV semi-amplitudes, \,$K_p$, and scaled with the resulting t-value to map the detection.

\subsection{Retrievals}

After computing the log-likelihood and CCF (from equations \ref{eq:logL} and \ref{eq:CCF}, respectively) \texttt{STARSHIPS} can then conduct retrievals to converge on optimal model parameters. As explained in \cite{boucher_characterizing_2021}, the retrieval uses the Markov Chain Monte Carlo framework \textit{emcee} \citep{foreman-mackey_emcee_2013}. Retrievals were done combining post-transit,  nights 3 and 5,  using 64 walkers. Free parameters for both atmospheric models are shown in Table \ref{tab:parameters}. We have put a prior on $K_{p}$ and $v_{rad}$ based on preliminary results, to anchor solutions. 


\section{Results} \label{sec:Results}

We applied the CCF and t-test to our WASP-33b data sets using models from our retrievals containing only H$_2$O, Fe, OH, TiO, VO, FeH or CO. Abundances are shown in Table \ref{tab:parameters}. Only CO was consistently detected, and results were similar for both models as explained in Section \ref{sec:Reduction}. We present the results from the model with a Guillot profile \citep{guillot_radiative_2010} containing only CO. Specifically, in Figure \ref{fig:combined-results} CO shows a non-detection in absorption pre-transit and a strong detection in absorption post-transit with an additional tentative blue-shifted inverted (emission) signal. 

\subsection{CO in absorption}\label{subsec:res-co}

The combined results of CO in absorption for both pre- and post-transit are shown in Figure \ref{fig:combined-results}. There is a strong absorption signal post-transit. The signal is present in both nights 3 and 5 of the post-transit observations. Individual nights are shown in Appendix \ref{sec:apendix_sepnights}. In pre-transit observations, however, we see no evidence of CO in absorption.

\subsection {CO in emission}

CO in emission is tentatively seen pre-transit as the dayside is shown blue-shifted. The strongest detection is shown on night two at $\sim$3$\sigma$ as seen in Figure \ref{fig:Idividual_nights}. Nights 1 and 4 do not show any strong signs of the emission signal, and thus, the combined plot in Figure \ref{fig:combined-results} shows no sign of a strong detection. Individual nights are shown in Appendix \ref{sec:apendix_sepnights}.

In our post-transit epochs, a tentative CO signal in emission is present (darker peak to the left of the brighter peak in Figure \ref{fig:combined-results}), with a blueshift of about 20\,km/s.

    \begin{figure*}[htbp]
    \centering
    \begin{minipage}{0.48\textwidth}
        \centering
        \includegraphics[width=\textwidth]{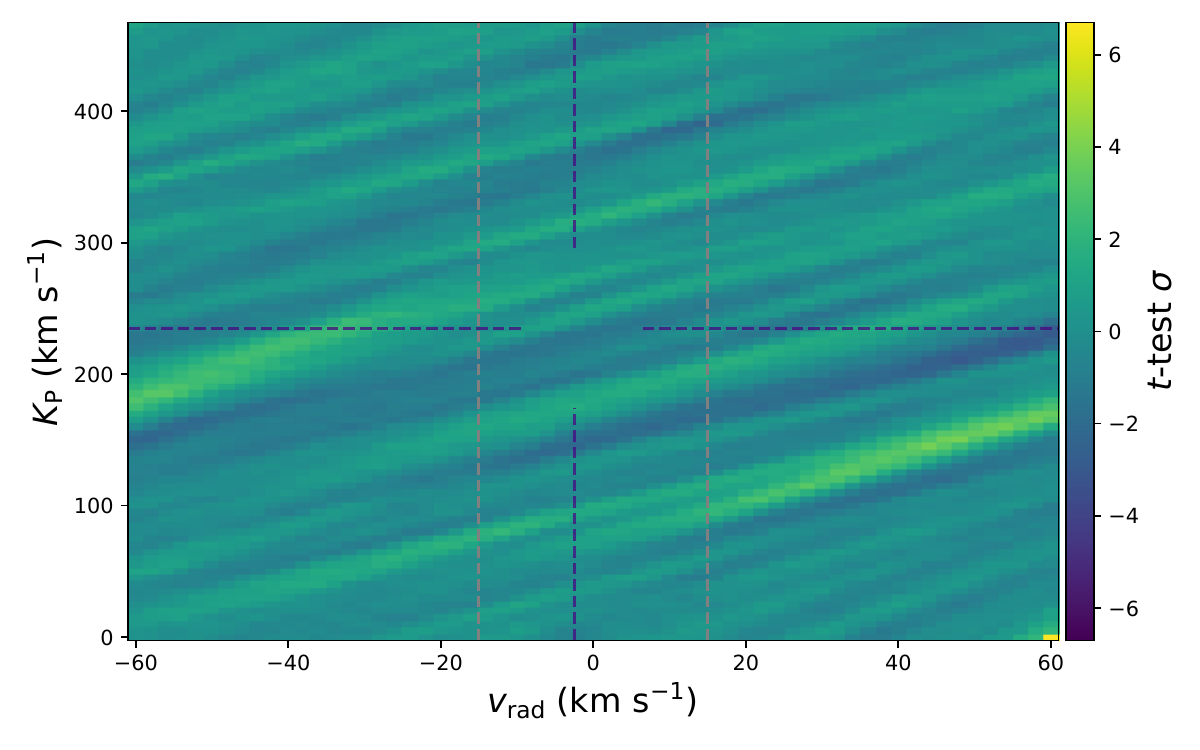}
    \end{minipage}\hfill
    \begin{minipage}{0.48\textwidth}
        \centering
        \includegraphics[width=\textwidth]{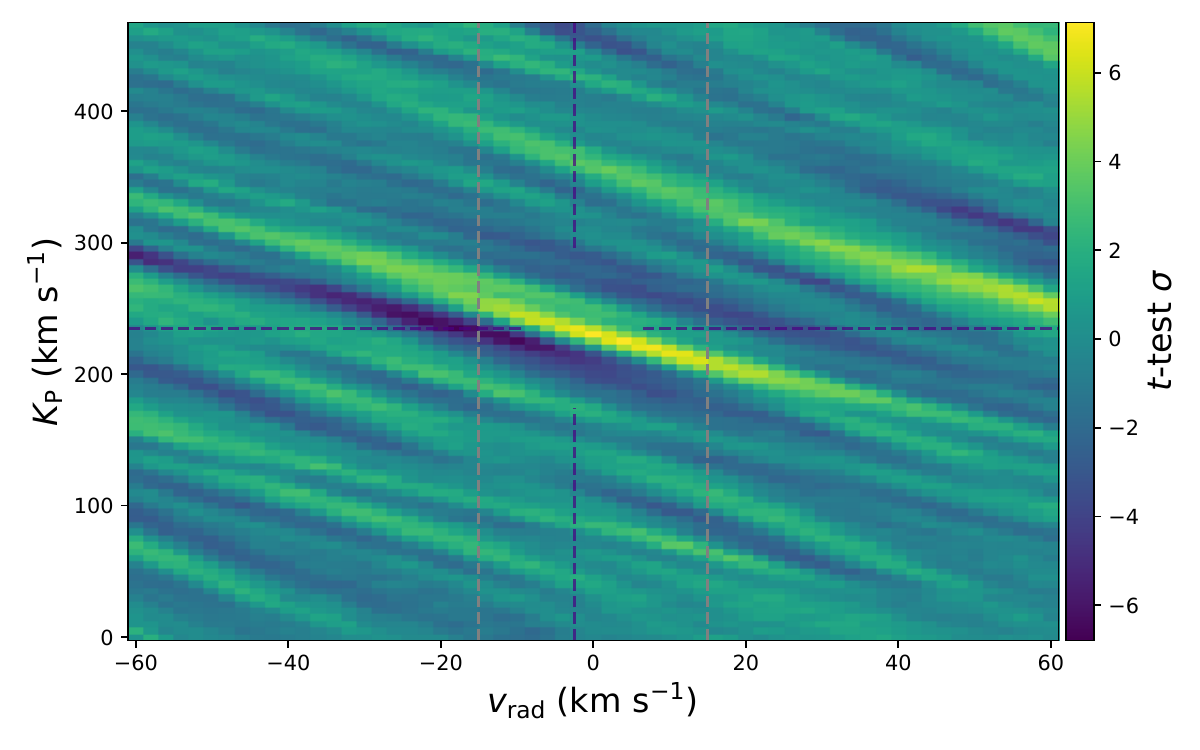}
    \end{minipage}
    \begin{minipage}{0.48\textwidth}
        \centering
        \includegraphics[width=\textwidth]{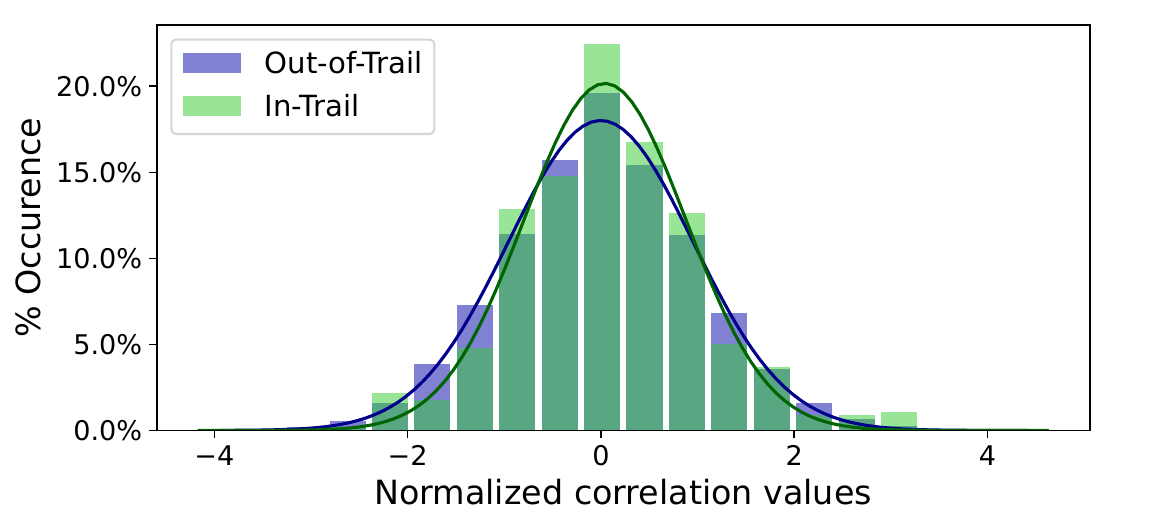}
    \end{minipage}
    \begin{minipage}{0.48\textwidth}
        \centering
        \includegraphics[width=\textwidth]{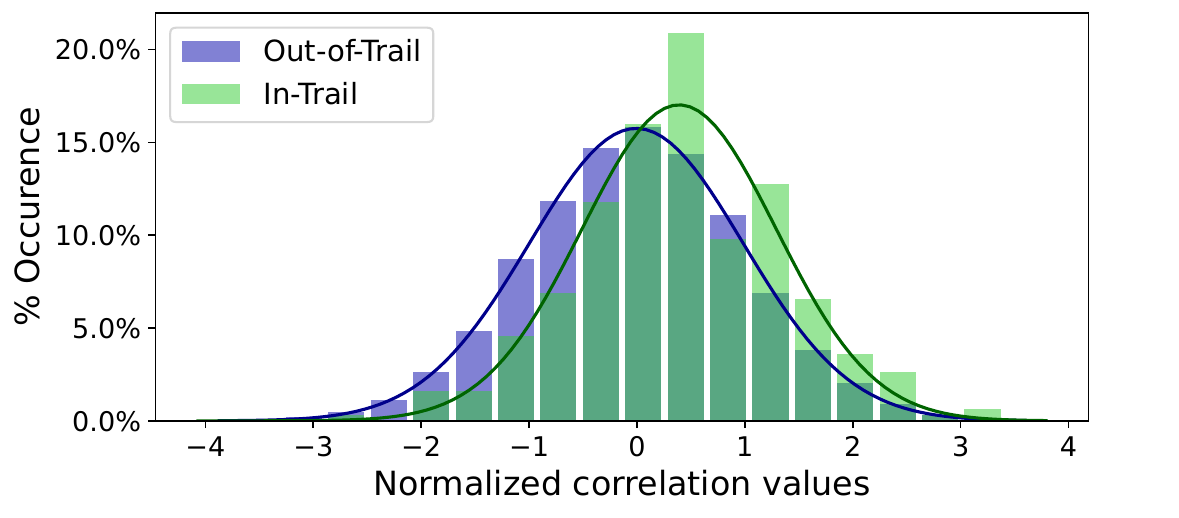}
    \end{minipage}

    \caption{ \emph{Top:} $K_p$ maps with scaled sigma. Grey vertical lines show the boundaries of in-trail data for t-test. Black vertical and horizontal dashed lines show peak absorption of post-transit data. \emph{Bottom:} T-test maps \emph{Left:} Pre-transit observations (Nights 1, 2, 4) combined. Sigma scaled by grid. No clear indication of CO in absorption. \emph{Right:} Post-transit observations (Nights 3 and 5). Sigma scaled to region $K_p<140 km/s$. Detection of CO in absorption at $K_p=230.00 km/s$ with a blue shift of -3.18 at about 6.6 $\sigma$. }
    \label{fig:combined-results}
\end{figure*}

\subsection{Retrievals Results}

We utilized the high-resolution retrievals present in the \texttt{STARSHIPS} code. The retrievals reveal a detection of CO in absorption post-transit, shown in Figure \ref{fig:retrievals}. Our priors and abundances are shown in Table \ref{tab:parameters} as well as in Appendix \ref{sec:apendix_retrieval}. Retrievals for both temperature profiles are shown. CO in absorption was retrieved from both models with a similar abundance.
Our best-fit model of CO using the Modified Guillot profile showed CO in both emission and absorption simultaneously. Regardless, a significant detection of CO in absorption was obtained (over 4$\sigma$). This simultaneous emission and absorption model may result from fitting a 3D planet, which exhibits both emission and absorption features from different regions, with a 1D atmospheric model.

\begin{table}[htbp]
    \centering
    \caption{MCMC Retrieval Parameters Priors and Results}
    \label{tab:parameters}
    \begin{tabularx}{\columnwidth}{lXXl}
        \toprule
        \textbf{Parameter} & \textbf{Prior} & \textbf{Results} & \textbf{Unit} \\
        \midrule
        \multicolumn{4}{c}{\textbf{Guillot Profile}} \\
        \midrule
        CO & $\mathcal{U}(-12, -0.5)$ & $-4.87^{+1.34}_{-0.36}$ & \\
        H$^{-}$ & $\mathcal{U}(-12, -0.5)$ & $-2.97^{+2.47}_{-5.18}$ & \\
        $T_p$ & $\mathcal{U}(400, 4500)$ & $3624.56^{+873.98}_{-1309.51}$ & K \\
        $K_p$ & $\mathcal{U}(200, 250)$ & $221.99^{+7.43}_{-7.73}$ & km/s \\
        $v_{\mathrm{rad}}$ & $\mathcal{U}(-20, 20)$ & $6^{+5}_{-5}$ & km/s \\
        $T_{p\gamma}$ & $\mathcal{U}(-2, 6)$ & $-0.68^{+3.10}_{-1.32}$ & \\
        $T_{p\kappa}$ & $\mathcal{U}(-3, 3)$ & $0.96^{+0.81}_{-0.86}$ & \\
        \midrule
        \multicolumn{4}{c}{\textbf{Modified Guillot Profile}} \\
        \midrule
        CO & $\mathcal{U}(-12, -0.5)$ & $-3.30^{+1.67}_{-1.68}$ & \\
        H\textsubscript{2}O & $\mathcal{U}(-12, -0.5)$ & $<-4.23$ & \\
        Fe & $\mathcal{U}(-12, -0.5)$ & $<-1.30$ & \\
        OH & $\mathcal{U}(-12, -0.5)$ & $<-2.24$ & \\
        VO & $\mathcal{U}(-12, -0.5)$ & $<-2.12$ & \\
        TiO & $\mathcal{U}(-12, -0.5)$ & $<-3.90$ & \\
        FeH & $\mathcal{U}(-12, -0.5)$ & $<-1.39$ & \\
        H$^{-}$ & $\mathcal{U}(-12, -0.5)$ & $-5.26^{+3.96}_{-2.92}$ & \\
        $T_p$ & $\mathcal{U}(400, 4500)$ & $1177.45^{+1333.88}_{-776.74}$ & K \\
        $K_p$ & $\mathcal{U}(200, 250)$ & $232.79^{+9.21}_{-9.56}$ & km/s \\
        $v_{\mathrm{rad}}$ & $\mathcal{U}(-10, 10)$ & $-2.24^{+9.63}_{-7.98}$ & km/s \\
        $\log_{10}\delta$ & $\mathcal{U}(-8, -3)$ & $-3.93^{+0.93}_{-0.85}$ & \\
        $\log_{10}\gamma$ & $\mathcal{U}(-2, 6)$ & $-0.56^{+1.84}_{-1.43}$ & \\
        $\alpha$ & $\mathcal{U}(-1, 1)$ & $0.66^{+0.35}_{-0.66}$ & \\
        $\log_{10}P_{\mathrm{trans}}$ & $\mathcal{U}(-8, 3)$ & $-3.56^{+4.59}_{-1.75}$ & \\
        \bottomrule     
    \end{tabularx}
    \begin{tablenotes}
        \small
        \item Note: The values are given with either their $\pm$ 1-$\sigma$ errors or their 2-$\sigma$ upper limits.
    \end{tablenotes}
\end{table}

\begin{figure}[htbp]
    \centering
    \includegraphics[width=0.45\textwidth]{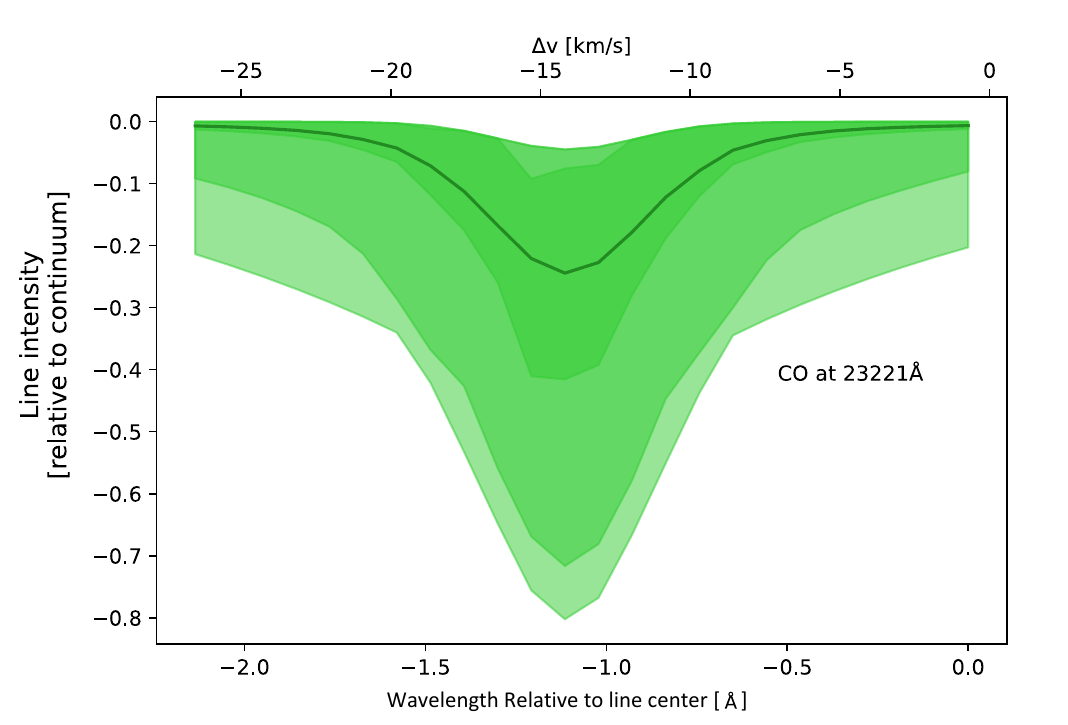}
    \includegraphics[width=0.45\textwidth]{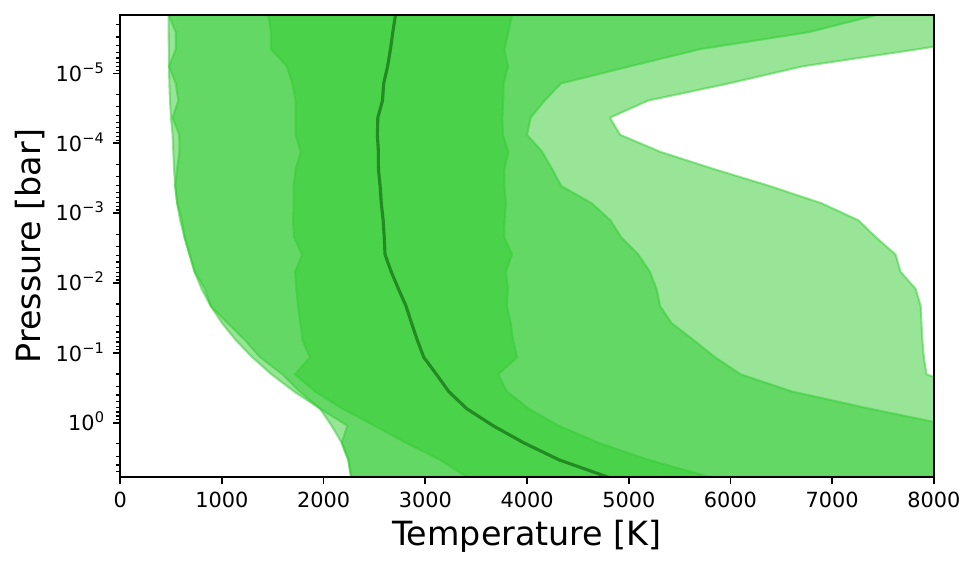}  
    \caption{\emph{Top:} Average line shape profile of CO from converged MCMC spectral retrievals. The black line shows the median of a selection of samples. The dark-shaded region represents their 1-$\sigma$, and the light-shaded region represents 2-$\sigma$ with the lightest region representing 3-$\sigma$ uncertainties. \emph{Bottom:} The TP profile where the shaded regions represent the same sigma values as the line profile.}
    \label{fig:retrievals}
    
\end{figure}

\section{Discussion} \label{sec:discussions}

We detect CO in absorption post-transit (evening) on the nightside of WASP-33b at $\sim$6.6$\sigma$. The detection appears to be accompanied by a tentative blue-shifted emission signal. We hypothesize that the blue-shifted CO in emission is from the dayside of the planet rotating into view. The planet's rotational velocity could induce shifts of up to  $\pm 7 \, \text{km/s}$, not including winds. Hot Jupiters are expected to have eastward, super-rotating winds, which on UHJs reach mangitudes of up to 10 $\, \text{km/s}$ \citep{wardenier_modelling_2023}. Strong winds could allow for a strong blue-shifted signal of the dayside coming from the limb of the planet. When additional retrievals were run, narrowing the prior of the {$v_{rad}$ to -40 to -5 km/s to encompass only the tentative emission feature, we retrieved the suspected CO in emission, indicating that this signal may be planetary in nature and not a reduction affect or an effect from the absorption CO signal. This blue-shifted CO emission, if confirmed by future observations, could be an additional confirmation of the thermally inverted dayside of WASP-33b. However, more detailed analysis is needed to confirm if this is planetary in nature. \cite{yang_evidence_2024} finds evidence of H$_2$O in emission on the nightside indicating a possible inverted temperature profile on its nightside. However, our CO absorption signal from the nightside of the planet confirms the expected non-inverted temperature-pressure profile. 

\cite{matthews_doppler_2024} found a tentative detection of CO absorption in the nightside of the non-transiting hot Jupiter HD 179949b. They combined spectra from phases 0.8 through 0.2. In contrast, we have chosen to separate our observations into pre- and post-transit epochs to probe the morning and evening terminators separately. 

In pre-transit observations, we see no detection of CO in absorption and tentative evidence of CO in emission. The emission was only shown in one out of the three nights at $\sim$3$\sigma$, making it tentative. We do not claim this as a detection. A non-detection of CO pre-transit is likely due to the evening terminator being warmer than the morning terminator. This configuration is expected due to heat transport via an eastward jet predicted by GCM simulations \citep{showman_atmospheric_2002} and observed in thermal phase curves \citep{cowan_model_2011,bell_comprehensive_2021} and in HRS by \cite{nortmann_crires_2024}, who observed that the morning terminator of WASP-127b is colder. \cite{yang_high-resolution_2023} found no evidence for significant day-to-night winds using transmission spectroscopy on WASP-33b, but those observations probe the atmosphere at lower pressures than the emission spectra presented here. Alternatively, the non-detection in pre-transit could be explained by a cloudy morning limb as seen in WASP-121b \citep{wardenier_modelling_2023}.

Based on our best-fit atmospheric models, we computed our cross-correlation using single-molecule absorption cross-sections. Models typically favored temperatures less than 2700 K, consistent with empirical measurements of nightside temperatures of ultra-hot Jupiters, including WASP-33b with the Spitzer Space Telescope \citep[1500 K][]{bell_comprehensive_2021,zhang_phase_2018-1}.

WASP-33 is a well known pulsating star \citep{herrero_wasp-33_2011,smith_thermal_2011}. The pulsations have been seen to affect high-resolution transit spectroscopy \citep{borsa_gaps_2021, cont_detection_2021,herman_dayside_2022}. However, these studies focused on lines that are also present in the stellar spectrum. CO should be fully dissociated in the hot stellar photosphere and therefore stellar pulsations should not affect our CO detection \citep{van_sluijs_carbon_2023}. Indeed, we see no evidence of periodic residuals affecting our CO detection. 

Our cross-correlation search for single molecules only robustly detected CO, but many other species including OH, H$_2$O, Fe, TiO, FeH and VO have been previously searched for in the atmosphere of WASP-33b. We therefore included these species in our retrieval exercise and obtained posteriors on some of their abundances, but we do not consider any of these to be robust detections since they were not detected in single-species cross-correlation searches. The results are shown in Appendix \ref{sec:Non-Detection}. The non-detection of these elements on the nightside of the planet could be due to multiple reasons, including an isothermal vertical profile at the relevant altitudes, their abundance may be below our sensitivity, or clouds could be masking certain regions.

 \section{Summary and Conclusions} \label{sec:conclusions}

We observed the ultra-hot Jupiter WASP-33b over 5 nights with SPIRou on the CFHT for a total of 14.25 hours. The observations included 3 pre-transit and 2 post-transit observations. Using the \texttt{STARSHIPS} open-source HRS analysis pipeline, we have made the strongest ground-based detection of nightside thermal emission from an exoplanet \citep[and only the third to date, cf.][]{matthews_doppler_2024, yang_evidence_2024}. This establishes ground-based high-resolution spectroscopy as a viable and cost-effective alternative to space-based full-orbit spectroscopic phase curves to characterize the nights-sides of exoplanets \citep[cf.][]{stevenson_deciphering_2014,mikal-evans_diurnal_2022}.

The detection of CO in absorption on the planet's nightside at about 6.6$\sigma$, combined with the numerous published detections of dayside emission features, confirms the expected vertical atmospheric structure: a temperature inversion on the irradiated dayside, and a non-inverted atmosphere on the nightside. The detection of CO in absorption on the evening terminator and absence of detection on the morning terminator is indicative of eastward heat transport via equatorial jet and is also consistent with the results of \cite{nortmann_crires_2024} for WASP-127b. The simultaneous presence of both the nightside absorption and hints of the dayside emission feature highlight that even single-epoch emission spectroscopy is sensitive to varied regions of the exoplanet.

\section*{Acknowledgements}

The results use observations made at the Canada-France-Hawaii telescope (CFHT), which is operated from Maunakea by the National Research Council of Canada, the Institut National des Sciences de l'univers of the Centre National de la Recherche Scientifique of France, and the University of Hawaii. We thank Etienne Artigau and Neil Cook for their work in Pre-Processing the data used in this paper from the CFHT. We additionally thank Romain Allart, Joost Wardenier, Vigneshwaran Krishnamurthy, and Vincent Yariv for their helpful discussions. This research was enabled in part by support provided by \textit{Calcul Québec} (\url{https://www.calculquebec.ca/en/}) and the Digital Research Alliance of Canada (\url{https://alliancecan.ca}). NBC acknowledges support from an NSERC Discovery Grant, a Tier 2 Canada Research Chair, and an Arthur B.\ McDonald Fellowship. The authors also thank the Trottier Space Institute and l’Institut de recherche sur les exoplanètes for their financial support and dynamic intellectual environment.


%

\vspace{5mm}
\facilities{
\begin{itemize}
    \item CFHT (SPIRou)
    \item \texttt{Exoplanet Archive}
\end{itemize}
}

\software{
    \begin{itemize}
        \item \texttt{astropy} \citep{astropy_collaboration_astropy_2022}
        \item \texttt{ipython}  \citep{perez_ipython_2007}
        \item \texttt{matplotlib} \citep{hunter_matplotlib_2007}
        \item \texttt{numpy} \citep{harris_array_2020}
        \item \texttt{scipy} \citep{virtanen_scipy_2020}
        \item \texttt{ExoFile} (Darveau-Bernier et al., in prep. \\ \url{https://github.com/AntoineDarveau/exofile})
        \item \texttt{pygtc} \citep{bocquet_pygtc_2016}
        \item \texttt APERO \citep{cook_apero_2022}
    \end{itemize}
    }





\appendix

\section{Observation Systematics}
\label{sec:systematics}

Airmass and signal-to-noise ratio of all observations. We used all exposures in our analysis, as removing the lowest signal-to-noise exposures did not affect our results. Observations are split into pre- and post-transit observations in Figure \ref{fig:system}. 

\begin{figure*}[htbp]

    \centering
    \includegraphics[width=1\textwidth]{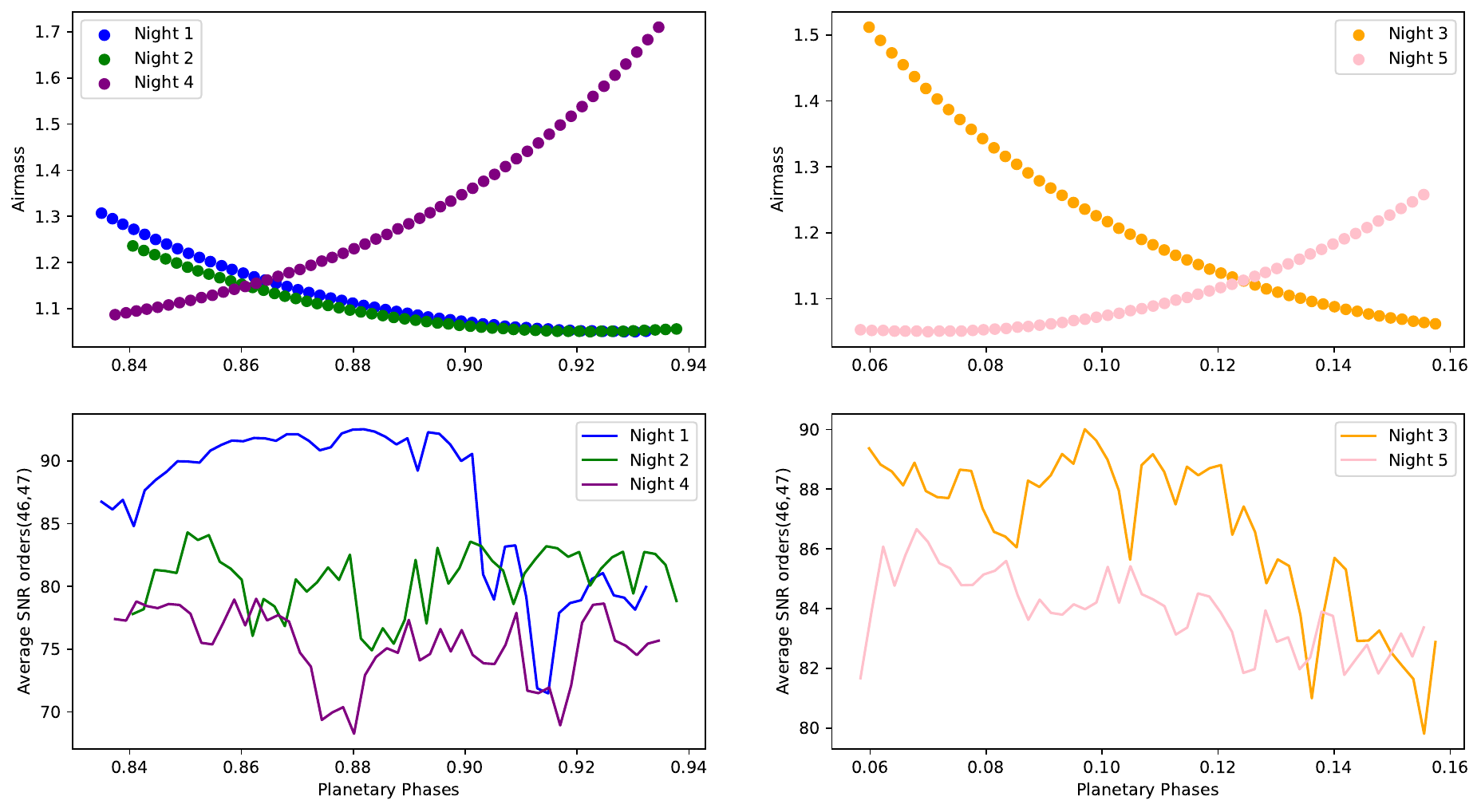} 
    \caption{Colors match observations depicted in Figure \ref{fig:obs_time}. Airmass and SNR of combined orders 46 and 47 are shown.}
    \label{fig:system} 
    
\end{figure*}

\section{Reduction Details}
\label{sec:reduction_details}

Step by step reduction method as presented in Section. \ref{sec:Reduction}. 
\begin{figure*}[htbp]

    \centering
   \includegraphics[width=1\textwidth]{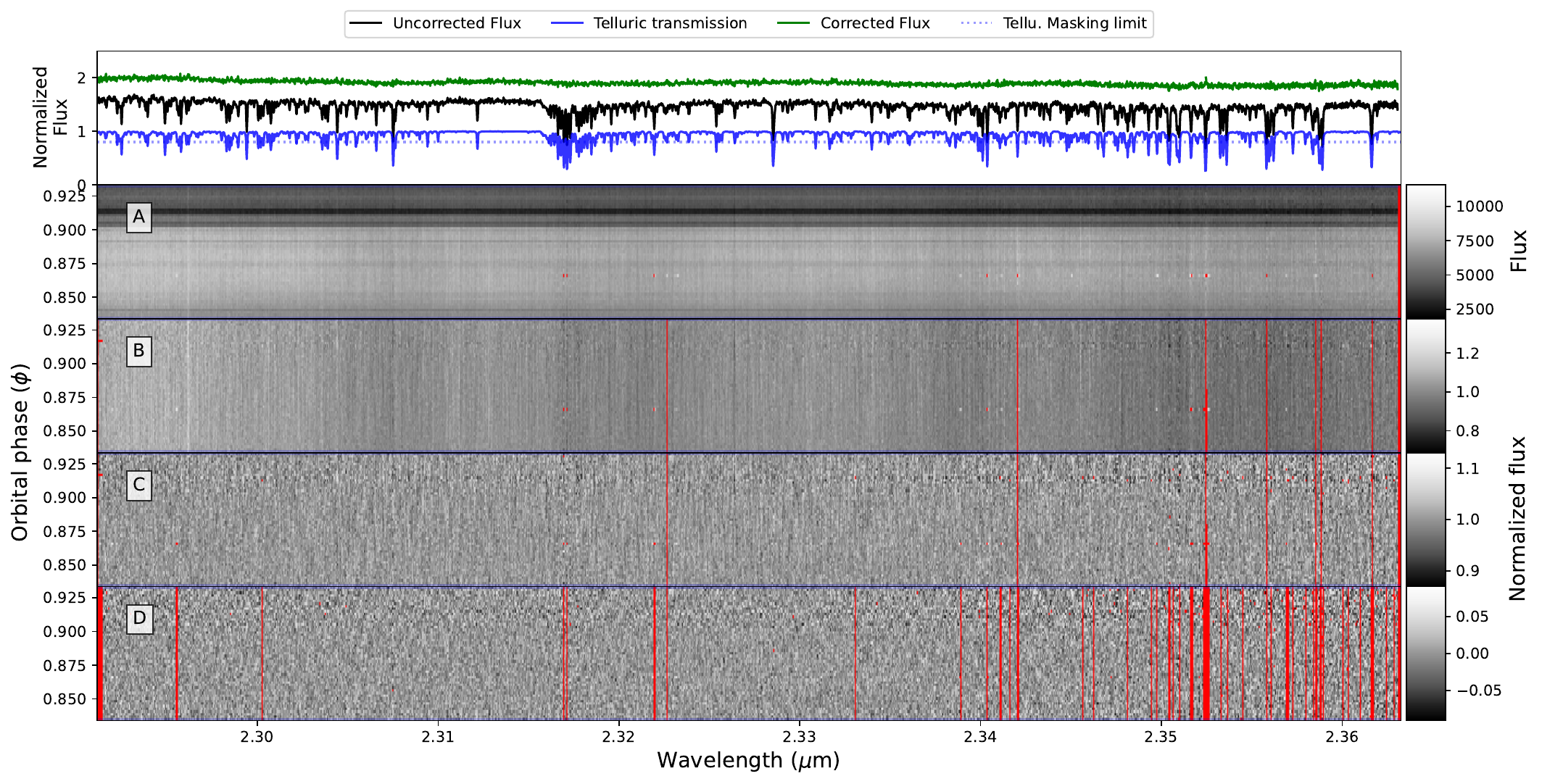} 
    \caption{Analysis steps that were applied to observed SPIRou spectral time series observations using \texttt{STARSHIPS}. Order 46 of Night 1 is shown. The top panel shows the uncorrected (black), the telluric-corrected (green) spectra, and the reconstructed telluric emission spectrum (blue). \textbf{Panel (A)}: Spectra corrected for tellurics with pixels masked by \texttt{APERO} shown in red. \textbf{Panel (B):} Spectra shifted into the stellar rest frame and normalized to the continuum level of the stellar reference spectrum. Most of the bad pixels were corrected or masked. \textbf{Panel (C):} Emission spectra divided by the reference spectrum \textbf{ Panel (D):} Final Emission spectra, principal component analysis (2 PC used) to correct for vertical residuals (caused by quasi-static residuals from either telluric contamination or the star).}
    
    \label{fig:red46} 
\end{figure*}

\section{Individual Nights}
\label{sec:apendix_sepnights}

Individual nights are shown for both pre- and post-transit. Pre-transit, no night shows evidence of CO in absorption post-transit, both nights exhibit CO in absorption at a detection significance of over 4$\sigma$. 

\begin{figure}[htbp]
    \centering
    \includegraphics[width=0.75\textwidth]{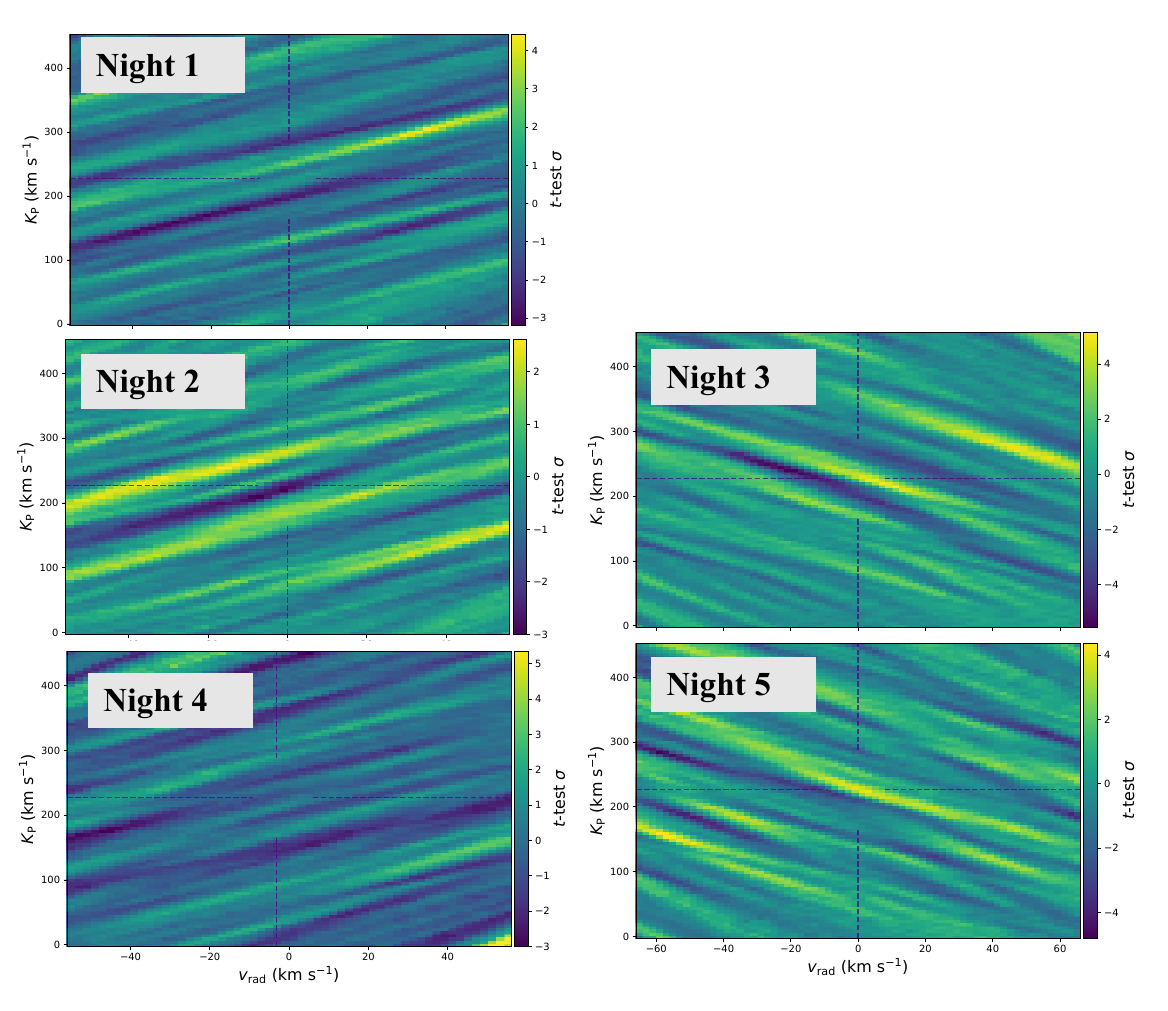}
   \caption{ Individual nights shown of CO. \emph{Left:} Pre-transit (Nights 1,2,5) sigma scaled by grid. \emph{Right:} Post-transit (Nights 3 and 5). Sigma scaled to $K_p > 350 km/s$}
    \label{fig:Idividual_nights}
\end{figure}

\section{Retrieval Posteriors}
\label{sec:apendix_retrieval}

Corner plots displaying the retrieval of both Guillot and Modified Guillot temperature profiles. Priors are listed in Table \ref{tab:parameters}. Both models are in close agreement for the abundance of CO. Additional elements only implemented into the Modified Guillot profile tested are shown below. We do not consider these robust detections, and the Kp Vsys maps are shown in Appendix \ref{sec:Non-Detection}. 

\begin{figure*}[htbp]
    \centering
    \includegraphics[width=0.7\textwidth]{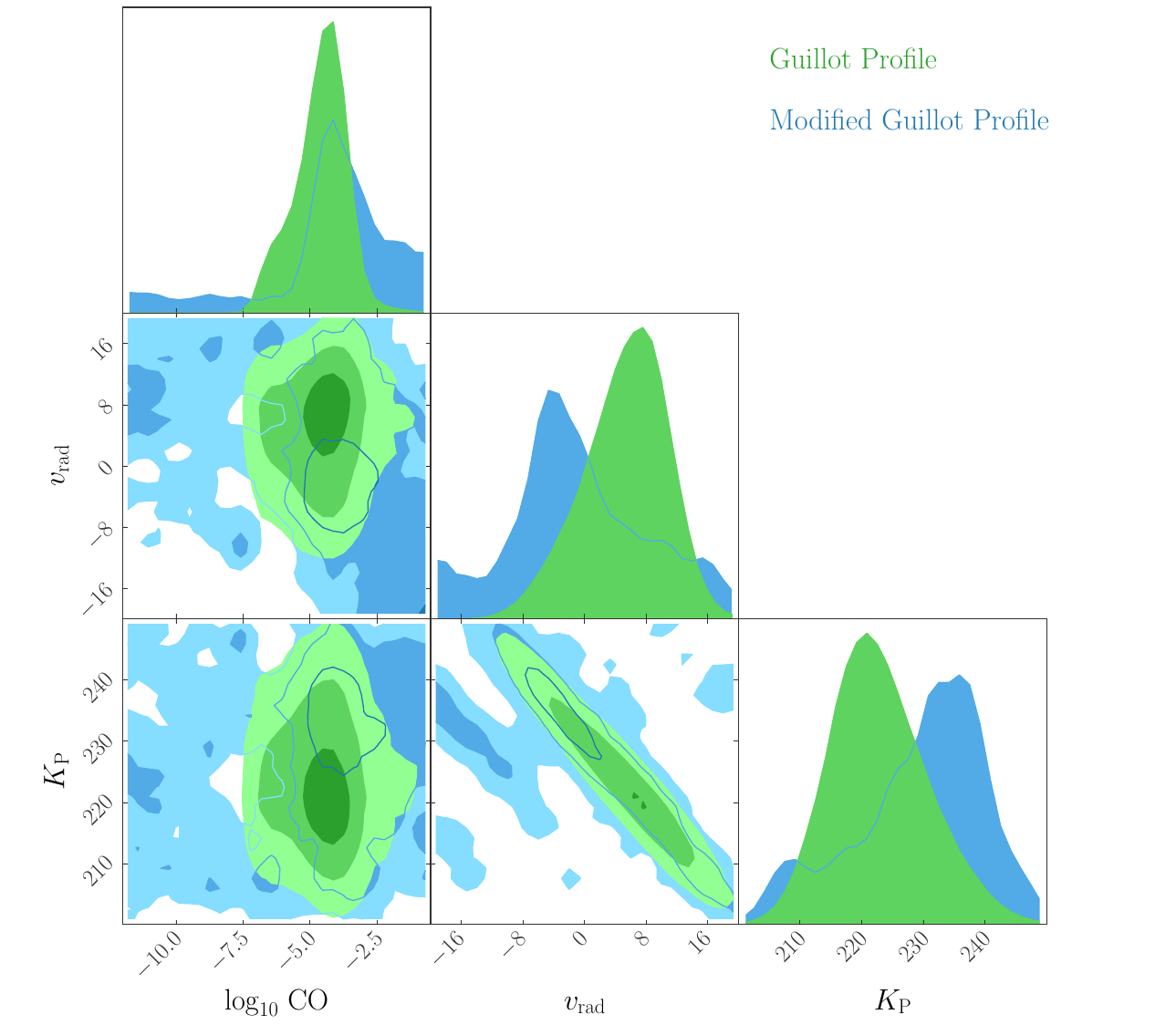}
    \vspace{1cm} 
    
    \begin{minipage}{0.3\textwidth}
        \centering
        \includegraphics[width=\textwidth]{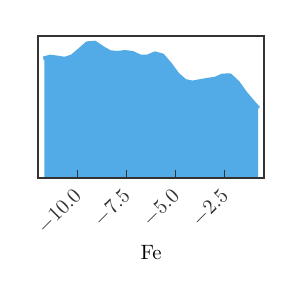}
    \end{minipage}
    \hfill
    \begin{minipage}{0.3\textwidth}
        \centering
        \includegraphics[width=\textwidth]{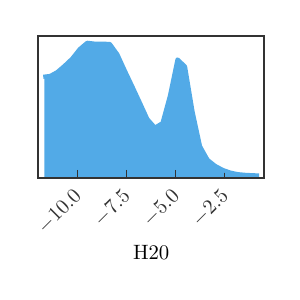}
    \end{minipage}
    \hfill
    \begin{minipage}{0.3\textwidth}
        \centering
        \includegraphics[width=\textwidth]{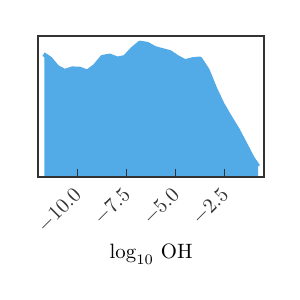}
    \end{minipage}
    \hfill
    \begin{minipage}{0.3\textwidth}
        \includegraphics[width=\textwidth]{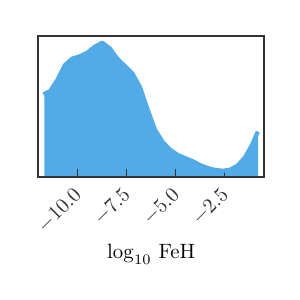}
    \end{minipage}
    \hfill
    \begin{minipage}{0.3\textwidth}
        \centering
        \includegraphics[width=\textwidth]{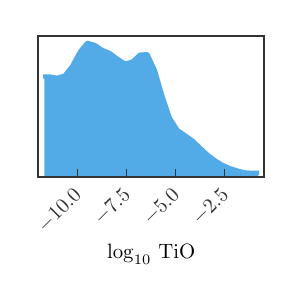}
    \end{minipage}
    \hfill
    \begin{minipage}{0.3\textwidth}
        \centering
        \includegraphics[width=\textwidth]{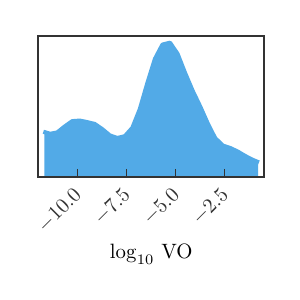}
    \end{minipage}
    
    \caption{Constraints on the atmospheric parameters retrieved post-transit (Nights 3 and 5 combined). The corner plot shows the posterior probability distributions from the MCMC retrievals. Modified Guillot profile with CO, H$_{2}$O, OH, Fe, TiO, FeH, and VO (Blue) and Guillot profile containing just contributions from CO (Green). }
    \label{fig:corner_plots}
\end{figure*}

\section{Other Species}
\label{sec:Non-Detection}

Additional molecules were implemented only in the Modified Guillot retrieval analysis, and best-fit models were individually cross-correlated using the same method as the CO detection. All orders were used in the cross-correlation as no difference was seen to the detections when limiting the orders.

We do not robustly detect any species other than CO. However, TiO may show a tentative detection pre-transit, but more observations would be needed to confirm this. We present the observations as combined pre-transit and combined post-transit data to compare with the CO detection. CO in molecular absorption is detected post-transit at greater than 4.5 $\sigma$ with the Modified Guillot Profile.

\begin{figure}[htbp]
    \centering
    \includegraphics[width=0.75\textwidth]{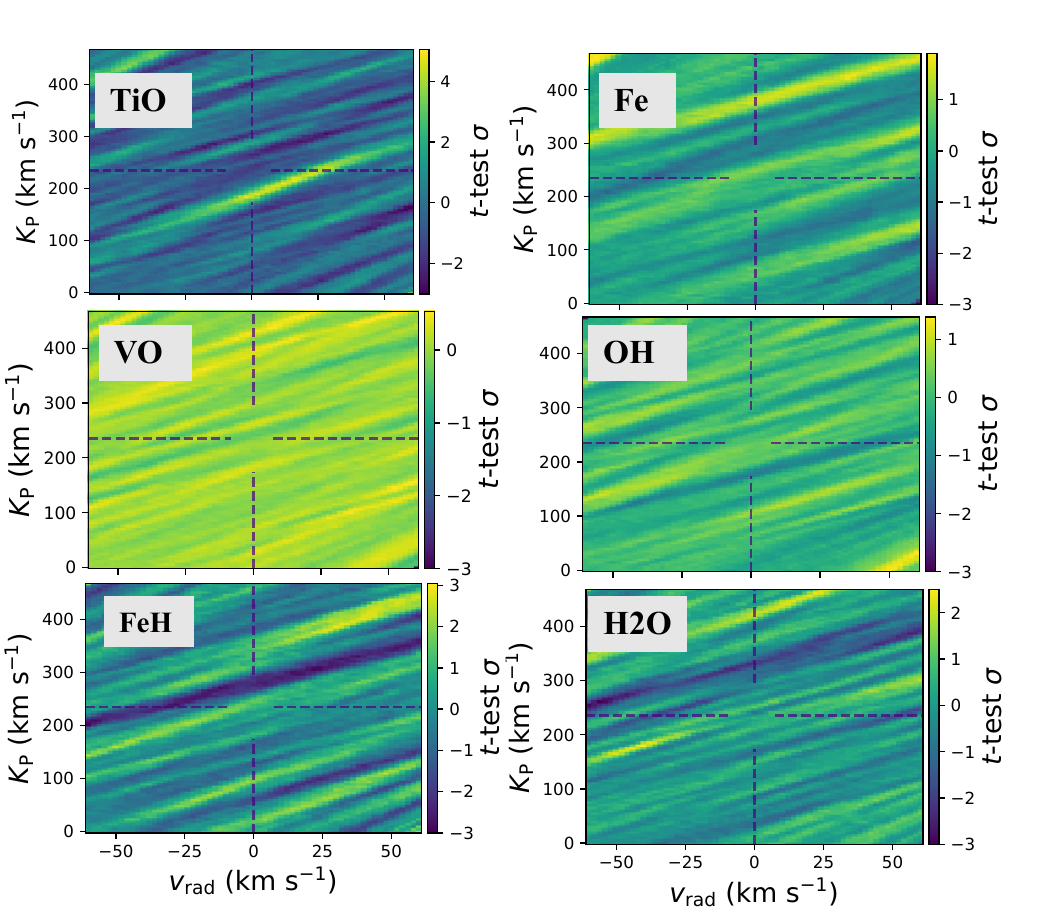}
   \caption{Other species used in the Modfied Guillot retrieval. Combined pre-transit observations (Nights 1, 2, and 4). Sigma is scaled by the grid. No species show any definitive detection. Each species is labeled and the dotted lines show the expected signal.}
    \label{fig:prettrans_mol}
\end{figure}

\begin{figure}[htbp]
    \centering
    \includegraphics[width=0.75\textwidth]{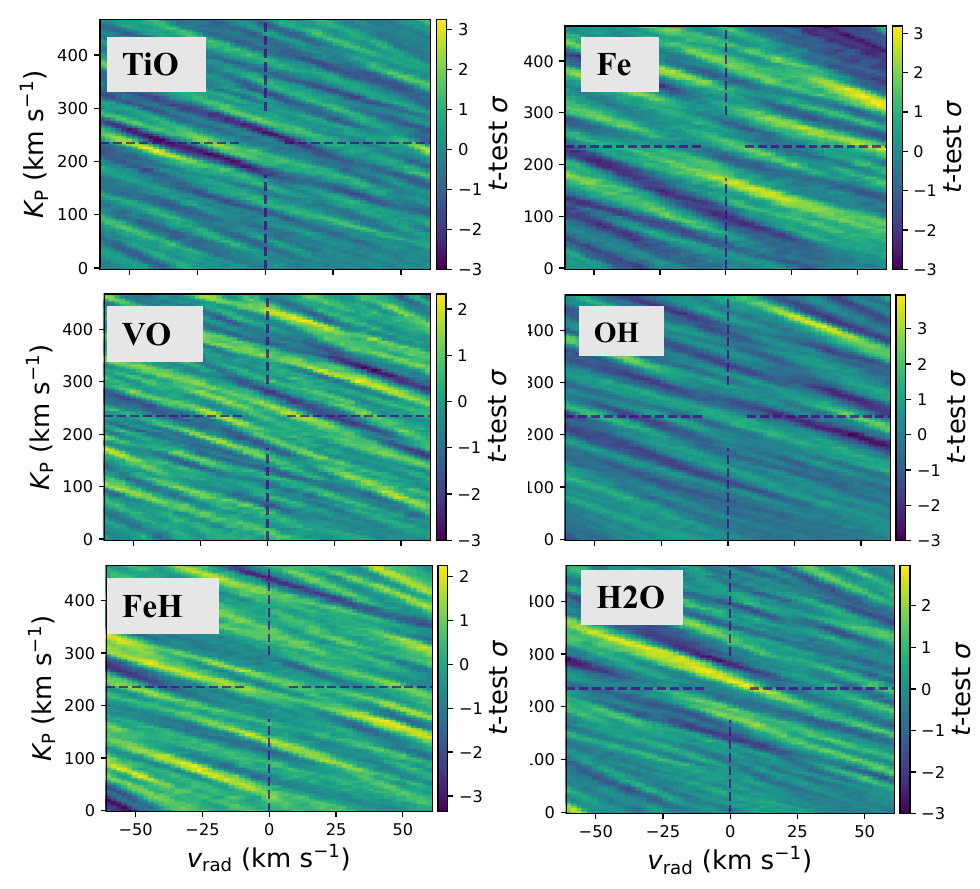}
   \caption{Same as figure \ref{fig:prettrans_mol} except combined post-transit observations (Nights 3 and 5). We do not consider any robust detections. Sigma is scaled by the grid. Each species is labeled, the dotted lines show the expected signal.}
    \label{fig:posttrans_mol}
\end{figure}






\bibliography{test}
\bibliographystyle{aasjournal}



\end{document}